\documentclass[aps,10pt, twocolumn]{revtex4}

\usepackage{amsmath,url,multirow}
\usepackage{graphicx}
\usepackage{amssymb}
\usepackage{epstopdf}
\begin{document}

\newtheorem{theorem}{Theorem}
\newtheorem{Remark}{Remark}
\newtheorem{lemma}{Lemma}
\newtheorem{cor}{Corollary } 
\newtheorem{proof}{Proof }

\title{On the effect of multiplicative noise in a supercritical pitchfork bifurcation}
\author{St. Reimann}
\affiliation{
Department of Management, Technology and Economics, ETH Zurich,
  Kreuzplatz 5, CH-8032 Zurich, Switzerland}
\begin{abstract}
The most important characteristic of {\em multiplicative noise} is that its effects of system's dynamics depends on the recent system's state. Consideration of multiplicative noise on self-referential systems including biological and economical systems therefore is of importance. In this note we study an elementary example. While in a deterministic super critical pitchfork bifurcation with positive bifurcation parameter $\lambda$ the positive branch $\sqrt{\lambda}$ is stable, multiplicative white noise $\lambda_t ={\lambda} + \sigma \zeta_t$ on the unique parameter reduces stability in that the system's state tends  to $0$ almost surely, even for ${\lambda}>0$, while for 'small' noise $\sigma < \sqrt{2 \lambda}$ the point $\sqrt{\lambda-\sigma^2/2}$ is a meta-stable state. In this case, correspondingly, the system will 'die out', i.e. $X_t \to 0$ within finite time. 
\end{abstract}

\maketitle
Natural systems exhibit a high degree of complexity due to being composed by a huge number of subsystems which interact in a strongly non-linear way. Under suitable conditions such systems can self-organize leading to a coherent behavior of a macroscopic scale in time and/or space \cite{Sornette2004}. Mathematical bifurcation theory has proved that, under fairly mild conditions, there exists reduction processes, series expansions and changes of variables of the many microscopic equations such that, near a critical point, systems' behavior is described by only a small number of differential equations depending on only one control parameter. In other words, close to a critical point, due to an increase of synchronization of parts, high-dimensional dynamics of complex systems is reduced to low-dimensional containing only one control parameter being described by
$ 
\dot{x}(t) = f_{\lambda} (x(t))
$ 
where $\lambda$ is some parameter system's controlling system's dynamics. Moreover it has been found that dynamics can be classified into a small number of archetypes  \cite{Arnold1988, Thom1972}. Due to openness, natural systems are subject to environmental influences. Without aiming to model the dynamics of these influences in detail, we assume that these influences can be approximated by noise. The properties of random dynamical systems have received great interest, see Arnold \cite{Arnold2002}. In principle, there are two ways of introducing noise into this system: Either {\em add} noise $\zeta(t)$ to the system's state, leading to
$
\dot{x}(t) \; = \; f_\lambda(x(t)) \; + \; \zeta(t)
$
or let noise act on the control parameter leading to
\begin{equation}
\dot{x}(t) \; = \; f_{\lambda_t} (x(t))
\end{equation}
It is well known, see  Horsthemke and Lefever \cite{HorsthemkeLefever1984} and Schenzle and Brand \cite{SchenzleBrand1979} for a general discussion, that for non-linear systems the effects additive and multiplicative noise have, are fundamentally different. While the effect of additive noise does not depend of the state of the system, the effect of multiplicative noise is state-dependent. Natural systems in which the effect of noise on the system's dynamics does dependent on the recent state are auto-catalytic chemical reactions or growth processes in developmental biology as well as in economy including the dynamics of financial markets. More generally speaking: in each system whose dynamics shows some degree of self-referentiality, the effect of exogenous noise will depend on the recent system's state. If noise is multiplicative, 'new' phenomena can occur, i.e. the noisy system can exhibit behavior, which is qualitatively different from that of the deterministic system, a phenomenon that has been coined {\em Noise-induced Transitions}. \\

In this note, we are particularly concerned with studying the effect of multiplicative noise in a super-critical pitchfork bifurcation. 
The mathematical equation of motion for the order parameter $x$ of the system representing a super-critical bifurcations, is known to be \cite{Berge1984}
\begin{equation}\label{normalform}
\dot{x} = \lambda \: x - \: x^3\end{equation}
where the cubic term represents a non-linear feedback which tends to limit the amplitude of the order parameter $x$. If the control parameter crosses some critical value $\lambda^*=0$, the trivial fixed point $0$ looses its stability, while two branches of fixed points $x^*_\pm = \pm \sqrt{\lambda}$ become stable.

Our particular question here is, what does happen to the stability of stable branch if the parameter is not constant but follows a stationary random process according to
\begin{equation}
\lambda_t \; = \; \lambda \; + \; \sigma \: \zeta_t
\end{equation}
where $\lambda \ge 0, \sigma > 0$ and $\zeta_t \sim {\mathcal N}(0,1)$ is standard Gaussian White noise.  In the following we restrict ourselves to the non-negative reals, i.e. $X_t \in [0,\infty)$, by imposing an absorbing boundary in $0$. In this case, 
we obtain the {\sc Ito} SDE 
\begin{equation}\label{Ito_model}
dX_t \; = \; \underbrace{\Big[ \lambda X_t - X_t^3 \Big]}_{\mu(X_t)} \: dt \; + \; \underbrace{\sigma X_t}_{\sigma(X_t)} \: dW_t
\end{equation}
where $W(t)$ is the standard Wiener process. 
By the non-linear transformation for positive $X_t \to Y_t =  \frac{1}{\sigma} \ln (X_t) $ the Ito process is transformed into $dY_t = \tilde{\mu}(Y_t) dt + dW_t$, where the transformed drift yields $\tilde{\mu}(Y_t) =\frac{1}{\sigma}\left[ \left(\lambda - \frac{\sigma^2}{2} \right) - Y_t^2 \right]$. The corresponding FPE yields, describing the diffusion of a particle in the potential $U(y)$ is
\begin{equation}
\partial_t \: \varphi(y,t) \; = \; \partial_y \Big( U'(y) \varphi(y,t) \Big) \; + \; \partial^2_u \; \varphi(y,t) 
\end{equation}
where, by introducing the critical noise strength $\sigma_* = \sqrt{2 \lambda}$ 
corresponding potential $U(y) = - \int^y \; \tilde{\mu}(z) dz$
yields 
\begin{equation}
U(y) = -\frac{1}{2\sigma} \; y \; \Big[\Big(\sigma_*^2 - \sigma^2\Big) - \frac{2}{3} \: y^2 \Big] 
\end{equation}

\begin{figure}[htbp] 

\includegraphics[width=6cm]{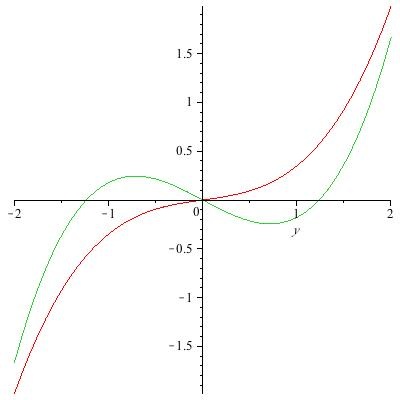}
   \caption{The Fokker-Planck potential $U(y)$ for $\sigma < \sigma^*$ (green) and $\sigma > \sigma^*$ (red)}
   \label{fig:U}

\end{figure}

From its shape, see Fig  \ref{fig:U}, it is apparent that the multiplicatively perturbed system behaves significantly different than the deterministic system. More precisely: $Y_t \to -\infty$ almost surely. Recalling that the system's state is $X_t = e^{\sigma \: Y_t}$, one obtains the following\\

{\bf RESULT 1:} {\em Let $\lambda, \sigma>0$. Then $X_t \to 0$ almost surely.
Moreover, for sufficiently weak noise $\sigma<\sigma_*$ the noisy system exhibits a unique positive meta-stable state. 
}\\

This is in contrast to the deterministic case $\sigma = 0$, where the fixed point $x^*=0$ is unstable, while the positive fixed point $x^* = \sqrt{2\lambda}$ is stable. Moreover, in the presence of noise, the system's state reaches $0$ in finite time

\begin{figure}[h]
\setlength{\unitlength}{1cm}
\hskip -2.0cm
\begin{minipage}[t]{5.5cm}
	\begin{picture}(5.5,5.5)
     \includegraphics[width=8cm]{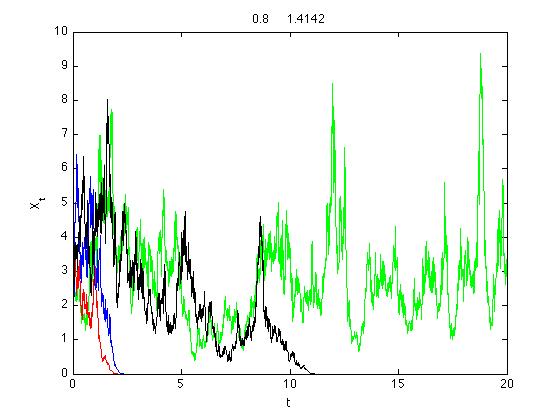}
	\end{picture} \par
	\parbox[t]{7.5cm}{Simulation of $X_t$ for the case $\sigma < \sigma^*$. Due to the existence of a meta-stable positive state trajectories have some finite life-time before they converge to $0$}
	\label{oscillator}
\end{minipage} \par
\hskip -2.0cm
\begin{minipage}[t]{5.5cm}
	\begin{picture}(5.5,5.5)
     \includegraphics[width=8cm]{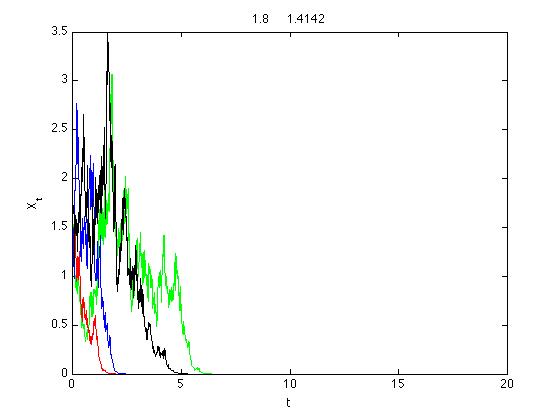}
	\end{picture} \par
	\parbox[t]{7.5cm}{Simulation of $X_t$ for the case $\sigma > \sigma^*$. Since the potential $U(y)$ is attracting to $-\infty$, $X_ \to 0$ rapidly. }
	\label{oscillator_rnd}
\end{minipage} 
\end{figure}

{\bf RESULT 2: }{\em Let $\lambda > 0$. For arbitrary small noise $\sigma>0$, the system will 'die', i.e. $X_t \to 0$, within finite time.\\

If multiplicative noise is sufficiently weak, i.e. $0< \sigma<\sigma_*$, there exists a meta-stable state in which the state $Y_t$ will be trapped for some time before escaping to $\infty$. The corresponding mean escape time from a finite state, can be approximated by the Arrhenius formula \cite{Gardiner1985}. 
While for $\sigma<\sigma_*$ the potential $U(y)$ is antisymmetric with respect to $0$ and has two extrema in $y^*_\pm = \pm \frac{1}{\sqrt{2}}\sqrt{\sigma_*^2 - \sigma^2}$, the maximal potential well to be crossed has hight
$\Delta U := U(y^*_-) - U(y^*_+) = 2 U(y_-) = \frac{\sqrt{2}}{3 \sigma} \left( \sigma_*^2 - \sigma^2 \right)^\frac{3}{2}$. Introducing $\zeta^2 := \sigma_*^2 - \sigma^2$, we obtain that the mean escape time approximately follows a stretched exponential function in $\zeta := \sqrt{\sigma_*^2 - \sigma^2}  > 0$
\begin{equation}\label{eqn:T}
T(y\to -\infty)  \; \sim\;  e^{\frac{\zeta^3}{\sigma}}.
\end{equation}
which is decreasing in $\sigma$! \\
 
We considered the noisy control parameter $\lambda_t = \lambda + \sigma \: \zeta_t$, where $\lambda > 0$ and $\zeta_t$ accounts for white noise. While in the deterministic case $\sigma = 0$, the positive branch is stable and the zero fixed point is unstable, the behavior of the parametrically perturbed system is significantly different: even for arbitrary small noise $\sigma>0$, the system's state $X_t$ will vanish within finite time, see eqn \ref{eqn:T}.  For sufficiently weak noise, there exists a positive meta-stable state.

\end{document}